\documentclass[aps,prl,twocolumn,superscriptaddress,linenumbers,showpacs,floatfix]{revtex4}

\usepackage{graphicx}
\DeclareGraphicsExtensions{.eps,.ps,.eps.gz,.png,.gz}
\usepackage[colorlinks,breaklinks]{hyperref}
\usepackage{dcolumn}
\usepackage{bm}
\usepackage{subfigure}
\usepackage{amsmath}
\usepackage{mathrsfs}
\usepackage{slashed}
\usepackage{eso-pic} 
\usepackage{color}
\usepackage{ifthen}
\usepackage{xspace}
\usepackage{multirow}
\usepackage{comment}
\usepackage{url}

\newcommand{\MET}{\ensuremath{\slashed E_{T}}}

\newcommand{\pt}{p_{T}}

\begin{document}





\title{Measurement of \boldmath $R = {\mathcal{B}\left( t \rightarrow 
Wb \right)/\mathcal{B}
\left(t \rightarrow Wq \right)} $ in Top--Quark--Pair Decays using 
Dilepton Events and the Full CDF Run II Data Set} 

\affiliation{Institute of Physics, Academia Sinica, Taipei, Taiwan 11529, Republic of China}
\affiliation{Argonne National Laboratory, Argonne, Illinois 60439, USA}
\affiliation{University of Athens, 157 71 Athens, Greece}
\affiliation{Institut de Fisica d'Altes Energies, ICREA, Universitat Autonoma de Barcelona, E-08193, Bellaterra (Barcelona), Spain}
\affiliation{Baylor University, Waco, Texas 76798, USA}
\affiliation{Istituto Nazionale di Fisica Nucleare Bologna, \ensuremath{^{ii}}University of Bologna, I-40127 Bologna, Italy}
\affiliation{University of California, Davis, Davis, California 95616, USA}
\affiliation{University of California, Los Angeles, Los Angeles, California 90024, USA}
\affiliation{Instituto de Fisica de Cantabria, CSIC-University of Cantabria, 39005 Santander, Spain}
\affiliation{Carnegie Mellon University, Pittsburgh, Pennsylvania 15213, USA}
\affiliation{Enrico Fermi Institute, University of Chicago, Chicago, Illinois 60637, USA}
\affiliation{Comenius University, 842 48 Bratislava, Slovakia; Institute of Experimental Physics, 040 01 Kosice, Slovakia}
\affiliation{Joint Institute for Nuclear Research, RU-141980 Dubna, Russia}
\affiliation{Duke University, Durham, North Carolina 27708, USA}
\affiliation{Fermi National Accelerator Laboratory, Batavia, Illinois 60510, USA}
\affiliation{University of Florida, Gainesville, Florida 32611, USA}
\affiliation{Laboratori Nazionali di Frascati, Istituto Nazionale di Fisica Nucleare, I-00044 Frascati, Italy}
\affiliation{University of Geneva, CH-1211 Geneva 4, Switzerland}
\affiliation{Glasgow University, Glasgow G12 8QQ, United Kingdom}
\affiliation{Harvard University, Cambridge, Massachusetts 02138, USA}
\affiliation{Division of High Energy Physics, Department of Physics, University of Helsinki, FIN-00014, Helsinki, Finland; Helsinki Institute of Physics, FIN-00014, Helsinki, Finland}
\affiliation{University of Illinois, Urbana, Illinois 61801, USA}
\affiliation{The Johns Hopkins University, Baltimore, Maryland 21218, USA}
\affiliation{Institut f\"{u}r Experimentelle Kernphysik, Karlsruhe Institute of Technology, D-76131 Karlsruhe, Germany}
\affiliation{Center for High Energy Physics: Kyungpook National University, Daegu 702-701, Korea; Seoul National University, Seoul 151-742, Korea; Sungkyunkwan University, Suwon 440-746, Korea; Korea Institute of Science and Technology Information, Daejeon 305-806, Korea; Chonnam National University, Gwangju 500-757, Korea; Chonbuk National University, Jeonju 561-756, Korea; Ewha Womans University, Seoul, 120-750, Korea}
\affiliation{Ernest Orlando Lawrence Berkeley National Laboratory, Berkeley, California 94720, USA}
\affiliation{University of Liverpool, Liverpool L69 7ZE, United Kingdom}
\affiliation{University College London, London WC1E 6BT, United Kingdom}
\affiliation{Centro de Investigaciones Energeticas Medioambientales y Tecnologicas, E-28040 Madrid, Spain}
\affiliation{Massachusetts Institute of Technology, Cambridge, Massachusetts 02139, USA}
\affiliation{University of Michigan, Ann Arbor, Michigan 48109, USA}
\affiliation{Michigan State University, East Lansing, Michigan 48824, USA}
\affiliation{Institution for Theoretical and Experimental Physics, ITEP, Moscow 117259, Russia}
\affiliation{University of New Mexico, Albuquerque, New Mexico 87131, USA}
\affiliation{The Ohio State University, Columbus, Ohio 43210, USA}
\affiliation{Okayama University, Okayama 700-8530, Japan}
\affiliation{Osaka City University, Osaka 558-8585, Japan}
\affiliation{University of Oxford, Oxford OX1 3RH, United Kingdom}
\affiliation{Istituto Nazionale di Fisica Nucleare, Sezione di Padova, \ensuremath{^{jj}}University of Padova, I-35131 Padova, Italy}
\affiliation{University of Pennsylvania, Philadelphia, Pennsylvania 19104, USA}
\affiliation{Istituto Nazionale di Fisica Nucleare Pisa, \ensuremath{^{kk}}University of Pisa, \ensuremath{^{ll}}University of Siena, \ensuremath{^{mm}}Scuola Normale Superiore, I-56127 Pisa, Italy, \ensuremath{^{nn}}INFN Pavia, I-27100 Pavia, Italy, \ensuremath{^{oo}}University of Pavia, I-27100 Pavia, Italy}
\affiliation{University of Pittsburgh, Pittsburgh, Pennsylvania 15260, USA}
\affiliation{Purdue University, West Lafayette, Indiana 47907, USA}
\affiliation{University of Rochester, Rochester, New York 14627, USA}
\affiliation{The Rockefeller University, New York, New York 10065, USA}
\affiliation{Istituto Nazionale di Fisica Nucleare, Sezione di Roma 1, \ensuremath{^{pp}}Sapienza Universit\`{a} di Roma, I-00185 Roma, Italy}
\affiliation{Mitchell Institute for Fundamental Physics and Astronomy, Texas A\&M University, College Station, Texas 77843, USA}
\affiliation{Istituto Nazionale di Fisica Nucleare Trieste, \ensuremath{^{qq}}Gruppo Collegato di Udine, \ensuremath{^{rr}}University of Udine, I-33100 Udine, Italy, \ensuremath{^{ss}}University of Trieste, I-34127 Trieste, Italy}
\affiliation{University of Tsukuba, Tsukuba, Ibaraki 305, Japan}
\affiliation{Tufts University, Medford, Massachusetts 02155, USA}
\affiliation{University of Virginia, Charlottesville, Virginia 22906, USA}
\affiliation{Waseda University, Tokyo 169, Japan}
\affiliation{Wayne State University, Detroit, Michigan 48201, USA}
\affiliation{University of Wisconsin, Madison, Wisconsin 53706, USA}
\affiliation{Yale University, New Haven, Connecticut 06520, USA}

\author{T.~Aaltonen}
\affiliation{Division of High Energy Physics, Department of Physics, University of Helsinki, FIN-00014, Helsinki, Finland; Helsinki Institute of Physics, FIN-00014, Helsinki, Finland}
\author{S.~Amerio\ensuremath{^{jj}}}
\affiliation{Istituto Nazionale di Fisica Nucleare, Sezione di Padova, \ensuremath{^{jj}}University of Padova, I-35131 Padova, Italy}
\author{D.~Amidei}
\affiliation{University of Michigan, Ann Arbor, Michigan 48109, USA}
\author{A.~Anastassov\ensuremath{^{v}}}
\affiliation{Fermi National Accelerator Laboratory, Batavia, Illinois 60510, USA}
\author{A.~Annovi}
\affiliation{Laboratori Nazionali di Frascati, Istituto Nazionale di Fisica Nucleare, I-00044 Frascati, Italy}
\author{J.~Antos}
\affiliation{Comenius University, 842 48 Bratislava, Slovakia; Institute of Experimental Physics, 040 01 Kosice, Slovakia}
\author{G.~Apollinari}
\affiliation{Fermi National Accelerator Laboratory, Batavia, Illinois 60510, USA}
\author{J.A.~Appel}
\affiliation{Fermi National Accelerator Laboratory, Batavia, Illinois 60510, USA}
\author{T.~Arisawa}
\affiliation{Waseda University, Tokyo 169, Japan}
\author{A.~Artikov}
\affiliation{Joint Institute for Nuclear Research, RU-141980 Dubna, Russia}
% ...

\collaboration{CDF Collaboration}
\altaffiliation[With visitors from]{
\ensuremath{^{a}}University of British Columbia, Vancouver, BC V6T 1Z1, Canada,
\ensuremath{^{b}}Istituto Nazionale di Fisica Nucleare, Sezione di Cagliari, 09042 Monserrato (Cagliari), Italy,
\ensuremath{^{c}}University of California Irvine, Irvine, CA 92697, USA,
\ensuremath{^{d}}Institute of Physics, Academy of Sciences of the Czech Republic, 182~21, Czech Republic,
\ensuremath{^{e}}CERN, CH-1211 Geneva, Switzerland,
\ensuremath{^{f}}Cornell University, Ithaca, NY 14853, USA,
\ensuremath{^{g}}University of Cyprus, Nicosia CY-1678, Cyprus,
\ensuremath{^{h}}Office of Science, U.S. Department of Energy, Washington, DC 20585, USA,
\ensuremath{^{i}}University College Dublin, Dublin 4, Ireland,
\ensuremath{^{j}}ETH, 8092 Z\"{u}rich, Switzerland,
\ensuremath{^{k}}University of Fukui, Fukui City, Fukui Prefecture, Japan 910-0017,
\ensuremath{^{l}}Universidad Iberoamericana, Lomas de Santa Fe, M\'{e}xico, C.P. 01219, Distrito Federal,
\ensuremath{^{m}}University of Iowa, Iowa City, IA 52242, USA,
\ensuremath{^{n}}Kinki University, Higashi-Osaka City, Japan 577-8502,
\ensuremath{^{o}}Kansas State University, Manhattan, KS 66506, USA,
\ensuremath{^{p}}Brookhaven National Laboratory, Upton, NY 11973, USA,
\ensuremath{^{q}}Queen Mary, University of London, London, E1 4NS, United Kingdom,
\ensuremath{^{r}}University of Melbourne, Victoria 3010, Australia,
\ensuremath{^{s}}Muons, Inc., Batavia, IL 60510, USA,
\ensuremath{^{t}}Nagasaki Institute of Applied Science, Nagasaki 851-0193, Japan,
\ensuremath{^{u}}National Research Nuclear University, Moscow 115409, Russia,
\ensuremath{^{v}}Northwestern University, Evanston, IL 60208, USA,
\ensuremath{^{w}}University of Notre Dame, Notre Dame, IN 46556, USA,
\ensuremath{^{x}}Universidad de Oviedo, E-33007 Oviedo, Spain,
\ensuremath{^{y}}CNRS-IN2P3, Paris, F-75205 France,
\ensuremath{^{z}}Universidad Tecnica Federico Santa Maria, 110v Valparaiso, Chile,
\ensuremath{^{aa}}The University of Jordan, Amman 11942, Jordan,
\ensuremath{^{bb}}Universite catholique de Louvain, 1348 Louvain-La-Neuve, Belgium,
\ensuremath{^{cc}}University of Z\"{u}rich, 8006 Z\"{u}rich, Switzerland,
\ensuremath{^{dd}}Massachusetts General Hospital, Boston, MA 02114 USA,
\ensuremath{^{ee}}Harvard Medical School, Boston, MA 02114 USA,
\ensuremath{^{ff}}Hampton University, Hampton, VA 23668, USA,
\ensuremath{^{gg}}Los Alamos National Laboratory, Los Alamos, NM 87544, USA,
\ensuremath{^{hh}}Universit\`{a} degli Studi di Napoli Federico I, I-80138 Napoli, Italy
}
\noaffiliation

\begin{abstract}
\noindent
We present a measurement of the ratio of the top-quark 
branching fractions 
$R=\mathcal{B}(t\rightarrow  Wb)/\mathcal{B}(t\rightarrow Wq)$, where 
$q$ represents any quark flavor,
in events with two charged leptons, imbalance in total transverse 
energy, and at least two  jets. 
The measurement uses 
proton--antiproton collision data at center-of-mass energy 
1.96 TeV, corresponding to an integrated luminosity of 8.7~fb$^{-1}$ 
collected with the Collider Detector at Fermilab during Run II of the Tevatron. 
We measure $R$ to be $0.87 \pm 0.07$, and extract
the magnitude of the top-bottom quark coupling to be 
$\left|V_{tb}\right| = 0.93 \pm 0.04$, 
assuming three generations of quarks.
Under these assumptions, a lower limit of $|V_{tb}|>0.85 (0.87)$ at 95 
(90) \% 
credibility level is set.

\end{abstract}

\maketitle                                                                                                                                      
In the standard model (SM) of fundamental interactions,
the top--quark decay 
rate into a $W$
boson and a down-type quark $q$ ($q = d, s, b$) is proportional to 
$\left|V_{tq}\right|^{2}$, the squared 
element of the Cabibbo-Kobayashi-Maskawa (CKM)
matrix~\cite{CKM}. 
In the hypothesis of three generations and unitarity for that 3$\times$3
matrix, and using the existing constraints on $V_{ts}$ and $V_{td}$, 
the magnitude of the top--bottom quark coupling is
$\left |V_{tb}\right|=0.99915^{+0.00002}_{-0.00005}$~\cite{PDG, Vts}.
Under these assumptions, 
the ratio of the branching fractions
\begin{equation} 
R =\frac{\mathcal{B}(t\rightarrow Wb)}{\mathcal{B}(t\rightarrow Wq)},
\label{eq:Rratio}
\end{equation}
is indirectly determined by the knowledge of 
$|V_{ts}|$ and $|V_{td}|$~\cite{PDG}~as
\begin{equation}
R = 
\frac{\left|V_{tb}\right|^2}
{\left|V_{tb}\right|^2+\left|V_{ts}\right|^2+\left|V_{td}\right|^2} 
= 0.99830^{+0.00004}_{-0.00009},
\label{eq:Rratio2}
\end{equation}
implying that the
top--quark decays almost exclusively to the $Wb$ final state. 
A deviation from this prediction would be an 
indication of non-SM physics, suggesting, for example, the existence of a 
fourth quark generation~\cite{atwood}.

The branching ratio and
$\left|V_{tb}\right|$ in Eq.~\ref{eq:Rratio2}
can be determined by studying the
rate of decays of pair-produced top--quarks into different quark flavors. 
In this article we report the measurement of $R$ in the sample of
top--quark pairs decaying leptonically ($t\bar{t}\rightarrow 
W^{+}qW^{-}\bar{q}\rightarrow q\bar{q} \ell \ell\nu\bar{\nu}$).
This method was used in previous 
measurements of $R$ 
by the CDF~\cite{cdfmeas} and the D0~\cite{d0meas} collaborations at 
the Fermilab Tevatron proton--antiproton collider. In the 
channel involving two charged leptons in the final state (dilepton 
channel), D0 measured 
$R=0.86\pm 0.05$~\cite{d0meas}. 
Recently the CDF collaboration 
updated its measurement in the channel involving a charged 
lepton and jets obtaining 
$R=0.94\pm 0.09$~\cite{cdfrlj},
both consistent with SM expectations.

A direct measurement of $|V_{tb}|$ can be
obtained from  the single-top-quark production cross 
section~\cite{singletop},
which is proportional to $\left|V_{tb}\right|^{2}$. 
By contrast, the branching ratio measurement reported here, based on 
top-pair-production, determines the size of $|V_{tb}|$ relative to the 
other CKM matrix elements. 
While the single top measurement depends on the absolute 
cross section, the branching ratio measurement depends on the relative 
yields for 0, 1, or 2 top decays to a $b$-quark.
In this sense the two measurements are complementary and the measurement 
of $|V_{tb}|$ presented here is less dependent on either the 
uncertainty on the theoretical 
calculation of the top-quark production cross section 
or many experimental uncertainties associated with its 
measurements.

This analysis studies events with two charged leptons, either electron 
($e$) or muon ($\mu$),
two neutrinos, and two or more jets in the final state; we do not search 
for $\tau$ leptons.
We use the full Run II data set,
corresponding 
to an integrated luminosity of 8.7~fb$^{-1}$
collected with the CDF II detector~\cite{CDFII} at the Tevatron 
at center-of-mass energy $\sqrt{s}=1.96$ TeV.



The CDF II detector~\cite{CDFII} consists of a particle
spectrometer embedded in a magnetic field of 1.4 T, with inner tracking 
chambers surrounded by  
electromagnetic and hadronic calorimeters segmented into towers
projecting to the interaction point, and outer muon detectors. 
A tracking system composed of a silicon microstrip detector
located at radial distance $r$ from the beam 1.5~$\leq 
r \leq$ 28~cm and of 
a drift chamber at 43 $\leq r \leq$ 132~cm, provides the 
reconstruction of charged-particle momentum and trajectories with 
full efficiency up to pseudorapidity $|\eta| \approx 1$~\cite{coord}.
The silicon microstrip detector is essential
for the detection of vertices displaced from the $p\bar{p}$ 
collision point signaling the decay of long-lived	
particles. 
A three-level, online event-selection system~\cite{trigger} is used 
to select events with an 
$e$ ($\mu$) candidate in the central detector region 
of pseudorapidity 
$| \eta | < 1.1 $,
with  $E_T$ ($p_{T}$) $>$ 18~GeV ($>$ 18~GeV/$c$), 
which form the data set for this analysis. 

The measurement of $R$ is based on the determination of the number of jets 
originated from $b$--quarks ($b$--jets) in
$t \bar t$ events reconstructed in  the dilepton final state. 
The dilepton signature consists of two high-$p_{T}$ charged 
leptons ($e$ or $\mu$),
large missing transverse energy $\slashed E_T$~\cite{coord} due to the 
undetected neutrinos from the leptonic $W$-boson decays,
and at least two hadronic jets. The identification of $b$-jets 
({\it tagging}) 
is performed by the {\sc secvtx} algorithm~\cite{secvtx}, 
which reconstructs secondary vertices separated from the primary 
collision vertex.

In order to better exploit the subsample-dependent signal-to-background 
ratio, we divide the  
sample into nine statistically independent subsamples according to 
dilepton flavor ($ee, \mu \mu, 
e\mu$) and $b$-tagging content (presence of 0,1, or 2 tags). 

As the number of $b$--jets in the event is related to the top--quark
branching fraction in the $Wb$ final state, we use
the number of observed and predicted 
events in the various subsamples as input to a likelihood 
function, which is maximized to extract $R$. 

The selection is similar to the one used by the CDF collaboration to 
measure 
the $t\bar{t}$  cross section in the dilepton channel~\cite{moon_prl}.
We select events with offline-reconstructed isolated oppositely-charged 
electrons ($E_{T}$ $\geq 20$~GeV) or
muons ($p_{T}$ $\geq 20$~GeV$/c$).
The contributions due to known standard model 
processes other than $t\bar{t}$ are further reduced by 
requiring a minimum \MET\ of $25$~GeV, increased to $50$~GeV if the 
direction of any lepton or jet is closer than 20$^{\circ}$ to the 
\MET\ direction, and \MET\ significance in excess of
4~(GeV)$^{1/2}$ ~\cite{moon_prl} for events with same-flavor
lepton pairs whose invariant mass is in a range of $\pm 15$~GeV/c$^2$ 
around the $Z$ boson mass~\cite{PDG}. 
Jets are reconstructed using a fixed-size cone algorithm~\cite{bhatti}, 
with radius of 0.4 in pseudorapidity-azimuthal angle $\eta-\phi$ space. 
We select events 
with at least two taggable~\cite{secvtx} jets with $E_{T}$ $\geq 
20$~GeV and $|\eta|<2$ after correcting for primary vertex position and jet 
energy scale.
Given the large size of the top--quark mass, we require
the sum of the transverse energies of the reconstructed  
leptons and jets, $H_T$, to be greater than 200~GeV.

The remaining background is composed of dibosons ({\it WW}, {\it WZ}, 
{\it ZZ}), Drell-Yan (DY) events 
($\tau^+\tau^-$, $e^+e^-$, $\mu^+ \mu^-$) with jets from 
initial (ISR) or final (FSR) state radiation and large \MET\ from energy 
mismeasurements, and associated production of $W$ bosons with multiple 
jets where 
one of the jets is misidentified as a charged lepton ({\it fakes}). 
The contributions of SM processes producing two real leptons are 
estimated using samples of events generated by Monte Carlo (MC) programs. 
The detector response is then simulated using 
a {\sc geant}~\cite{geant} based software package.
A combination of data and Monte Carlo samples is used to  estimate 
the contribution of jets misidentified as leptons~\cite{moon_prl}.
Diboson processes are simulated using {\sc pythia}~\cite{Pythia} and 
normalized to their next-to-leading order in strong interaction 
coupling cross sections,
$\sigma_{WW} $= 11.34 $\pm$ 0.68~pb, $\sigma_{WZ} $= 3.47 $\pm$ 0.21~pb, 
$\sigma_{ZZ} $= 3.62 $\pm$ 0.22~pb
\cite{campb}.
Drell-Yan and Z$\rightarrow \ell \ell$ events with associated jets are 
generated using 
{\sc alpgen}~\cite{alpgen}, with hadronization simulated using {\sc 
pythia}. 

Signal $t\bar{t}$  events are modeled using the {\sc 
powheg}~\cite{powheg}
generator, with hadronization simulated using {\sc pythia}.
A top--quark mass value of 
$172.5$~GeV/$c^2$, consistent with recent measurements~\cite{topmass}, is 
assumed. 


Due to the high purity of the $t\bar{t}$  signal in dilepton events, 
it is possible to perform a
measurement of the $t\bar{t}$ cross section in the sample without 
requiring $b$-tagging.
This result, free of any assumption on 
$\mathcal{B}(t\rightarrow Wb)$, is then used to predict the yield of 
top--quark 
events in the various tagging categories.
After the selection we find 286 events, which constitutes the {\it pretag} 
sample,
with an expected background of $54\pm7$  events. The 
largest background contributions are due to events containing jets 
misidentified 
as leptons and Drell-Yan events. From this we measure 
$\sigma_{p\bar{p} \rightarrow t\bar{t}} = 7.64 \pm 0.55\text{(stat)~pb},$
in agreement with previous results~\cite{moon_prl}.

In order to compare data and expectations in the nine subsamples 
we predict the amount of signal and background in each of them.
In those subsamples containing one or two $b$-tagged jets, we estimate the 
number 
of expected background events following the same 
strategy used in the $b$-tagged dilepton cross section 
measurement~\cite{moon_prl}. We use these estimates to calculate
the background in the subsamples with zero $b$-tags
by subtracting their sum from the total  
background in the pretag sample.
All background estimates are independent of $R$. A summary of SM 
expectations and observed events by tagging category is given in 
Table~\ref{tab:btagall}.

\begin{table}[!h]
	\centering
\renewcommand{\arraystretch}{0.5}

\begin{tabular}{l l l l}
%

\hline
\hline

\rule[-2mm]{0mm}{7mm} Process & Pretag & 1 tag& 2 tags	 \\\hline

\rule[-2mm]{0mm}{7mm}Dibosons &~5.4$\pm$0.6&0.66$\pm$0.10 & 0.035$\pm$0.014\\
\rule[-2mm]{0mm}{7mm}DY+LF   &10.7$\pm$1.6&1.50$\pm$0.70& 0.029$\pm$0.015\\
\rule[-2mm]{0mm}{7mm}DY+HF   & \ \ N/A&0.63$\pm$0.12& 
0.17$\pm$0.06\\
\rule[-2mm]{0mm}{7mm}Fakes   &21.8$\pm$4.3&5.6 $\pm$1.9&1.0 $\pm$0.5\\	 \hline		

\rule[-2mm]{0mm}{7mm}Total background   &~54$\pm$7&  8.3$\pm$ 2.1& 1.25$\pm$0.53\\\hline	
\rule[-2mm]{0mm}{7mm}\textit{$t\bar{t}$} ($\sigma$=7.4 pb) &223$\pm$20&100$\pm$ 9&29$\pm$4\\\hline	

\rule[-2mm]{0mm}{7mm}Total prediction  &278$\pm$21&110$\pm$10&30.8$\pm$4.2 \\

								
\rule[-2mm]{0mm}{7mm}Observed   &286&      96& 35	\\
\hline
\hline
\end{tabular}
\caption{
Summary of background contributions, $t\bar{t}$ SM
expectations (assuming $|V_{tb}|=1$), and data candidates by tagging 
categories for the 8.7 fb$^{-1}$ data sample. HF and LF indicate Heavy Flavor and Light Flavor jets.
\label{tab:btagall}
}
\end{table}

\begin{figure}[!htbp]
\centering
\includegraphics[width=.45\textwidth]{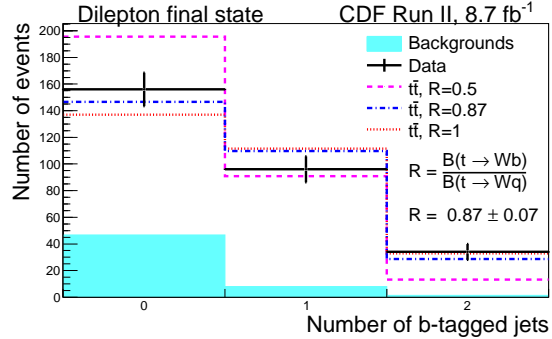}
\caption{
Number of events observed in data and expected for 
various values of $R$ as a function of identified $b$-jets.
\label{fig:dilep}
}
\end{figure}

The jet $b$-tagging efficency is measured in MC samples using the 
{\sc secvtx} algorithm after checking that the identified jet originates 
from the
hadronization of a bottom quark. This efficiency is corrected for differences
between data and simulation.
Mistagging occurs if jets from light-flavor quarks are mistakenly identified as 
coming from $b$--jets, and
its efficiency is calculated using
data templates and parametrized as a function of event variables
such as jet energy and number of tracks  in $\eta$ and $\pt$ intervals. 
In $t\bar{t}$ events we find an efficiency of $\simeq 40$ \% for tagging 
$b$-jets and a mistagging probability, of $\simeq 1$ \%.
Both efficiencies are used as inputs to the final fit. 
In the likelihood we include the possibility of reconstructing a third 
jet.
The number of $t\bar{t}$ signal events expected in each bin of the 
likelihood is a function of the probability 
for a jet to be tagged, which depends on $R$ 
since a $b$-quark-generated jet is more likely to be $b$-tagged.
In Fig.~\ref{fig:dilep} 
the number of events observed in data and expected for
different values of $R$ in the different tagging categories is shown.
The number of $t\bar{t}$ events expected in each bin is obtained by multiplying the
number of signal events before requiring $b$-tagging by the 
$R$-dependent probability of having 0, 1, or 2 $b$-tagged jets in the 
event.

In order to extract $R$ we maximize the likelihood

\begin{equation}
\textit{L} = \prod_i \ \mathscr{P}(\mu^i_\text{exp} \ (R, x_j) | N_\text{obs}^i)   
\prod_j G (x_i | \bar{x}_j, \sigma_j),
\end{equation}
where the index $i$ runs over the nine subsamples; 
$\mathscr{P}(\mu_\text{exp}^i \ (R, x_j) | N_\text{obs}^i)$ is the Poisson 
probability to 
observe $N_\text{obs}^i$ events, given the expected value $\mu_\text 
{exp}^i$;
and $G (x_i| \bar{x}_j, \sigma_j)$  are Gaussian probability density 
functions describing the knowledge 
of nuisance parameters $x_j$, with
mean $\bar{x}_j$ and standard deviation $\sigma_j$.
These nuisance parameters describe luminosity, background estimates, 
selection acceptances, and relevant efficiencies.
By using the same fit parameters for common sources of systematic 
uncertainties, correlations  
among different channels are taken into account. 

In the likelihood maximization $R$ is left as a free parameter.
In addition, we evaluate the effect of several contributions not accounted 
for among nuisance parameters.
We estimate the
systematic uncertainty due to imperfect modeling of
initial-state and final-state gluon radiation
by varying their amount
in simulated events~\cite{isrfsr} and taking as uncertainty the difference of the result with respect
to the nominal one.
The contribution from the jet-energy scale 
is estimated by varying its value by $\pm $1 standard 
deviation~\cite{isrfsr}, refitting the data, and taking as uncertainty the 
difference of the result with respect to the nominal result.
We find
\begin{equation}
R = 0.871 \pm 0.045 \text{\footnotesize{(stat)}}^{+0.058}_{-0.057}  
 \text{\footnotesize{(syst)}} 
 = 0.87 \pm 0.07. 
\end{equation}

To evaluate the effect of each nuisance parameter on the total systematic 
uncertainty, we 
perform the fit by individually fixing 
each nuisance parameter to a value corresponding to an excursion of 
one-standard deviation from its mean. The most important 
contributions to the $R$ systematic uncertainty are reported in 
Table~\ref{tab:systematics}.

\begin{table}[!h]
        \centering
                \begin{tabular}{lc}

                \hline
		\hline
Source & Syst. uncertainty\\
\hline
Correction to $b$-tagging efficiency & \\
in data and MC& +0.045,  -0.040\\
$\sigma_{t\bar{t}}$ & $\pm$ 0.01 \\ 
Luminosity & +0.009, -0.012 \\
Jet energy scale  & +0.033, -0.025 \\
ISR and FSR & +0.013, -0.025\\
\hline
Total systematic uncertainty & +0.059, -0.057 \\
Statistical uncertainty & $\pm $0.045\\
\hline
Total uncertainty & +0.074, -0.073 \\\hline
\hline
\end{tabular}
\caption{Systematic effects contributing the largest uncertainty to the measurement of $R$.
\label{tab:systematics}}
 \end{table}

To determine 
the credibility level limit on $R$ 
we follow a Bayesian statistical approach.
We use a uniform prior probability density for $R$ in the physical 
interval [0,1].
To obtain the 
posterior probability distribution for $R$, 
we integrate over all nuisance parameters using non-negative 
Gaussian distributions as prior probabilities. 
We obtain $R>0.73 (0.76)$ at 95 (90) \% credibility level.
From Eq.~(\ref{eq:Rratio2}) and the assumptions therein 
we obtain $|V_{tb}|=0.94\pm 0.04$ and $|V_{tb}|> 0.85 (0.87)$ at 95 (90) 
\% 
credibility level. 

In summary, in this Letter we present a measurement of the ratio of the top--quark 
branching fractions 
$R=\mathcal{B}(t\rightarrow  Wb)/\mathcal{B}(t\rightarrow Wq)$ 
in a sample of $t\bar{t}$ candidate events where bot $W$ bosons
from the top-quarks decay into leptons ($e$ or $\mu$). The $t\bar{t}$
are reconstructed 
using the CDFII 
detector from a dataset corresponding to 8.7~fb$^{-1}$ from $p\bar{p}$ 
collisions at $\sqrt{s}=$1.96~TeV.
The result, $R=0.87 \pm 0.07$, is consistent with previous 
measurements by CDF~\cite{cdfmeas} and D0~\cite{d0meas} collaborations and 
differs from the SM expectation by $\approx 1.8 \sigma$.


We thank the Fermilab staff and the technical staffs of the
participating institutions for their vital contributions. This work
was supported by the U.S. Department of Energy and National Science
Foundation; the Italian Istituto Nazionale di Fisica Nucleare; the
Ministry of Education, Culture, Sports, Science and Technology of
Japan; the Natural Sciences and Engineering Research Council of
Canada; the National Science Council of the Republic of China; the
Swiss National Science Foundation; the A.P. Sloan Foundation; the
Bundesministerium f\"ur Bildung und Forschung, Germany; the Korean
World Class University Program, the National Research Foundation of
Korea; the Science and Technology Facilities Council and the Royal
Society, United Kingdom; the Russian Foundation for Basic Research;
the Ministerio de Ciencia e Innovaci\'{o}n, and Programa
Consolider-Ingenio 2010, Spain; the Slovak R\&D Agency; the Academy
of Finland; the Australian Research Council (ARC); and the EU community
Marie Curie Fellowship Contract No. 302103.

\nocite{*}
\appendix


\begin{thebibliography}{150}
\bibitem{coord} We use a cylindrical coordinate system 
where the $z$ axis is along the proton beam direction, $\phi$ is the azimuthal angle, and 
$\theta$ is the polar angle. Pseudorapidity is $\eta = -\ln\tan(\theta/2)$, 
while transverse momentum is $p_{T}=\left|p\right|\sin\theta$, and transverse energy is $E_{T}=E\sin\theta$. 
Missing transverse energy, \MET, is defined as the magnitude of $-\sum_{i}E_{T}^{i}\hat{n}_{i}$, 
where $\hat{n}_{i}$ is the unit vector in the azimuthal plane 
that points from the beam line to the $i$th calorimeter tower.
\end{thebibliography}
\end{document}